\documentstyle[10pt,epsf,epsfig,dp_delphititle]{dp_delphi}
\makeindex
\def\DpPaperGroup{EP}
\def\DpPaperRef{2000-081}
\def\DpDate{7 June 2000}
\def\DpAuthors{DELPHI Collaboration}
\def\DpSubmit{(Physics Letters B490(2000)61)}
\def\DpTitle{{Rapidity-Rank Structure of 
\boldmath ${\mrm{p}\overline{\mrm{p}}}$~Pairs \\
 in Hadronic ${{\mrm{Z}^0}}$~Decays}}
\def\DpComment{ }
\def\DpEMail{ }

\newcommand{\mrm} {\mathrm}
\newcommand{\bfm} {\boldmath}

\newcommand{\into}{{\ifmmode \rightarrow \else $\rightarrow$\fi}}
\newcommand{\epem}{{\ifmmode e^+e^- \else $e^+e^-$\fi}}
\newcommand{\zz}{{\ifmmode Z^0~ \else $Z^0~$\fi}}
\newcommand{\sqs}{{\ifmmode \sqrt s~ \else $\sqrt s~$\fi}}
\newcommand{\de}{{\ifmmode ^{\circ} \else ${^\circ}$\fi}}
\newcommand{\p}{{\ifmmode p~ \else $p~$\fi}}
\newcommand{\bp}{{\ifmmode \bar{p}~ \else $\bar{p}~$\fi}}
\newcommand{\ps}{{$p$'s~}}
\newcommand{\bps}{{$\bar{p}$'s~}}
\newcommand{\qqb}{{\ifmmode q\,\bar{q}~ \else $q\,\bar{q}~$\fi}}
\newcommand{\ppb}{{\ifmmode p\,\bar{p}~ \else $p\,\bar{p}~$\fi}}
\newcommand{\ppbx}{{\ifmmode p\,\bar{p} \else $p\,\bar{p}$\fi}}
\newcommand{\pmpb}{{\ifmmode p\,M\,\bar{p}~ \else $p\,M\,\bar{p}~$\fi}}
\newcommand{\pmpbx}{{\ifmmode p\,M\,\bar{p} \else $p\,M\,\bar{p}$\fi}}
\newcommand{\dymin}{{\ifmmode \Delta y_{min}~\else$\Delta y_{min}~$\fi}}
\newcommand{\dyminx}{{\ifmmode \Delta y_{min}\else$\Delta y_{min}$\fi}}
\newcommand{\R}{{\ifmmode \cal{R}~ \else $\cal{R}~$\fi}}
\newcommand{\Rx}{{\ifmmode \cal{R} \else $\cal{R}$\fi}}
\newcommand{\Rdy}{{$\Rx(\dyminx)~$}}
\newcommand{\Rdyx}{{$\Rx(\dyminx)$}}
\newcommand{\llb}{{$\Lambda\,\bar{\Lambda}~$}}
\newcommand{\llbx}{{$\Lambda\,\bar{\Lambda}$}}
\newcommand{\BBb}{{\ifmmode B\,\bar{B}~ \else $B\,\bar{B}~$\fi}}
\newcommand{\BBbx}{{\ifmmode B\,\bar{B} \else $B\,\bar{B}$\fi}}
\newcommand{\dyppb}{{$\Delta y(p \bar{p})~$}}
\newcommand{\dyppbx}{{$\Delta y(p \bar{p})$}}

\begin{document}
\makeatletter
\newcount\@tempcntc
\def\@citex[#1]#2{\if@filesw\immediate\write\@auxout{\string\citation{#2}}\fi
  \@tempcnta\z@\@tempcntb\m@ne\def\@citea{}\@cite{\@for\@citeb:=#2\do
    {\@ifundefined
       {b@\@citeb}{\@citeo\@tempcntb\m@ne\@citea\def\@citea{,}{\bf ?}\@warning
       {Citation `\@citeb' on page \thepage \space undefined}}%
    {\setbox\z@\hbox{\global\@tempcntc0\csname b@\@citeb\endcsname\relax}%
     \ifnum\@tempcntc=\z@ \@citeo\@tempcntb\m@ne
       \@citea\def\@citea{,}\hbox{\csname b@\@citeb\endcsname}%
     \else
      \advance\@tempcntb\@ne
      \ifnum\@tempcntb=\@tempcntc
      \else\advance\@tempcntb\m@ne\@citeo
      \@tempcnta\@tempcntc\@tempcntb\@tempcntc\fi\fi}}\@citeo}{#1}}
\def\@citeo{\ifnum\@tempcnta>\@tempcntb\else\@citea\def\@citea{,}%
  \ifnum\@tempcnta=\@tempcntb\the\@tempcnta\else
   {\advance\@tempcnta\@ne\ifnum\@tempcnta=\@tempcntb \else \def\@citea{--}\fi
    \advance\@tempcnta\m@ne\the\@tempcnta\@citea\the\@tempcntb}\fi\fi}
 
\makeatother
\begin{titlepage}
\pagenumbering{roman}
\CERNpreprint{\DpPaperGroup}{\DpPaperRef} 
\date{{\small\DpDate}} 
\title{\DpTitle} 
\address{\DpAuthors} 
\begin{shortabs} 
\noindent
%
\noindent
The rapidity-rank structure of \ppb pairs is used to analyze the
mechanism of baryon production in hadronic \zz decay.
The relative occurrence of the rapidity-ordered configuration \pmpb,
where $M$ is a meson, and that of \ppb adjacent pairs is compared.
The data are found to be consistent with predictions from a mechanism
producing adjacent-rank \ppb pairs, without requiring `string-ordered'
\pmpb configurations.
An upper limit of 15\% at 90\% confidence is determined for the
\pmpb contribution.
\end{shortabs}
\vfill
\begin{center}
\DpSubmit \ \\ 
\DpComment \ \\
\DpEMail \ \\
\end{center}
\vfill
\clearpage
\headsep 10.0pt
\addtolength{\textheight}{10mm}
\addtolength{\footskip}{-5mm}
\begingroup
%
\newcommand{\DpName}[2]{\hbox{#1$^{\ref{#2}}$},\hfill}
\newcommand{\DpNameTwo}[3]{\hbox{#1$^{\ref{#2},\ref{#3}}$},\hfill}
\newcommand{\DpNameThree}[4]{\hbox{#1$^{\ref{#2},\ref{#3},\ref{#4}}$},\hfill}
\newskip\Bigfill \Bigfill = 0pt plus 1000fill
\newcommand{\DpNameLast}[2]{\hbox{#1$^{\ref{#2}}$}\hspace{\Bigfill}}
%
\footnotesize
\noindent
\DpName{P.Abreu}{LIP}
\DpName{W.Adam}{VIENNA}
\DpName{T.Adye}{RAL}
\DpName{P.Adzic}{DEMOKRITOS}
\DpName{I.Ajinenko}{SERPUKHOV}
\DpName{Z.Albrecht}{KARLSRUHE}
\DpName{T.Alderweireld}{AIM}
\DpName{G.D.Alekseev}{JINR}
\DpName{R.Alemany}{VALENCIA}
\DpName{T.Allmendinger}{KARLSRUHE}
% ...
\normalsize
\endgroup
\titlefoot{Department of Physics and Astronomy, Iowa State
     University, Ames IA 50011-3160, USA
    \label{AMES}}
\titlefoot{Physics Department, Univ. Instelling Antwerpen,
     Universiteitsplein 1, B-2610 Antwerpen, Belgium \\
     \indent~~and IIHE, ULB-VUB,
     Pleinlaan 2, B-1050 Brussels, Belgium \\
     \indent~~and Facult\'e des Sciences,
     Univ. de l'Etat Mons, Av. Maistriau 19, B-7000 Mons, Belgium
    \label{AIM}}
\titlefoot{Physics Laboratory, University of Athens, Solonos Str.
     104, GR-10680 Athens, Greece
    \label{ATHENS}}
\titlefoot{Department of Physics, University of Bergen,
     All\'egaten 55, NO-5007 Bergen, Norway
    \label{BERGEN}}
\titlefoot{Dipartimento di Fisica, Universit\`a di Bologna and INFN,
     Via Irnerio 46, IT-40126 Bologna, Italy
    \label{BOLOGNA}}
\titlefoot{Centro Brasileiro de Pesquisas F\'{\i}sicas, rua Xavier Sigaud 150,
     BR-22290 Rio de Janeiro, Brazil \\
     \indent~~and Depto. de F\'{\i}sica, Pont. Univ. Cat\'olica,
     C.P. 38071 BR-22453 Rio de Janeiro, Brazil \\
     \indent~~and Inst. de F\'{\i}sica, Univ. Estadual do Rio de Janeiro,
     rua S\~{a}o Francisco Xavier 524, Rio de Janeiro, Brazil
    \label{BRASIL}}
\titlefoot{Comenius University, Faculty of Mathematics and Physics,
     Mlynska Dolina, SK-84215 Bratislava, Slovakia
    \label{BRATISLAVA}}
\titlefoot{Coll\`ege de France, Lab. de Physique Corpusculaire, IN2P3-CNRS,
     FR-75231 Paris Cedex 05, France
    \label{CDF}}
\titlefoot{CERN, CH-1211 Geneva 23, Switzerland
    \label{CERN}}
\titlefoot{Institut de Recherches Subatomiques, IN2P3 - CNRS/ULP - BP20,
     FR-67037 Strasbourg Cedex, France
    \label{CRN}}
\titlefoot{Now at DESY-Zeuthen, Platanenallee 6, D-15735 Zeuthen, Germany
    \label{DESY}}
\titlefoot{Institute of Nuclear Physics, N.C.S.R. Demokritos,
     P.O. Box 60228, GR-15310 Athens, Greece
    \label{DEMOKRITOS}}
\titlefoot{FZU, Inst. of Phys. of the C.A.S. High Energy Physics Division,
     Na Slovance 2, CZ-180 40, Praha 8, Czech Republic
    \label{FZU}}
\titlefoot{Dipartimento di Fisica, Universit\`a di Genova and INFN,
     Via Dodecaneso 33, IT-16146 Genova, Italy
    \label{GENOVA}}
\titlefoot{Institut des Sciences Nucl\'eaires, IN2P3-CNRS, Universit\'e
     de Grenoble 1, FR-38026 Grenoble Cedex, France
    \label{GRENOBLE}}
\titlefoot{Helsinki Institute of Physics, HIP,
     P.O. Box 9, FI-00014 Helsinki, Finland
    \label{HELSINKI}}
\titlefoot{Joint Institute for Nuclear Research, Dubna, Head Post
     Office, P.O. Box 79, RU-101 000 Moscow, Russian Federation
    \label{JINR}}
\titlefoot{Institut f\"ur Experimentelle Kernphysik,
     Universit\"at Karlsruhe, Postfach 6980, DE-76128 Karlsruhe,
     Germany
    \label{KARLSRUHE}}
\titlefoot{Institute of Nuclear Physics and University of Mining and Metalurgy,
     Ul. Kawiory 26a, PL-30055 Krakow, Poland
    \label{KRAKOW}}
\titlefoot{Universit\'e de Paris-Sud, Lab. de l'Acc\'el\'erateur
     Lin\'eaire, IN2P3-CNRS, B\^{a}t. 200, FR-91405 Orsay Cedex, France
    \label{LAL}}
\titlefoot{School of Physics and Chemistry, University of Lancaster,
     Lancaster LA1 4YB, UK
    \label{LANCASTER}}
\titlefoot{LIP, IST, FCUL - Av. Elias Garcia, 14-$1^{o}$,
     PT-1000 Lisboa Codex, Portugal
    \label{LIP}}
\titlefoot{Department of Physics, University of Liverpool, P.O.
     Box 147, Liverpool L69 3BX, UK
    \label{LIVERPOOL}}
\titlefoot{LPNHE, IN2P3-CNRS, Univ.~Paris VI et VII, Tour 33 (RdC),
     4 place Jussieu, FR-75252 Paris Cedex 05, France
    \label{LPNHE}}
\titlefoot{Department of Physics, University of Lund,
     S\"olvegatan 14, SE-223 63 Lund, Sweden
    \label{LUND}}
\titlefoot{Universit\'e Claude Bernard de Lyon, IPNL, IN2P3-CNRS,
     FR-69622 Villeurbanne Cedex, France
    \label{LYON}}
\titlefoot{Univ. d'Aix - Marseille II - CPP, IN2P3-CNRS,
     FR-13288 Marseille Cedex 09, France
    \label{MARSEILLE}}
\titlefoot{Dipartimento di Fisica, Universit\`a di Milano and INFN-MILANO,
     Via Celoria 16, IT-20133 Milan, Italy
    \label{MILANO}}
\titlefoot{Dipartimento di Fisica, Univ. di Milano-Bicocca and
     INFN-MILANO, Piazza delle Scienze 2, IT-20126 Milan, Italy
    \label{MILANO2}}
\titlefoot{Niels Bohr Institute, Blegdamsvej 17,
     DK-2100 Copenhagen {\O}, Denmark
    \label{NBI}}
\titlefoot{IPNP of MFF, Charles Univ., Areal MFF,
     V Holesovickach 2, CZ-180 00, Praha 8, Czech Republic
    \label{NC}}
\titlefoot{NIKHEF, Postbus 41882, NL-1009 DB
     Amsterdam, The Netherlands
    \label{NIKHEF}}
\titlefoot{National Technical University, Physics Department,
     Zografou Campus, GR-15773 Athens, Greece
    \label{NTU-ATHENS}}
\titlefoot{Physics Department, University of Oslo, Blindern,
     NO-1000 Oslo 3, Norway
    \label{OSLO}}
\titlefoot{Dpto. Fisica, Univ. Oviedo, Avda. Calvo Sotelo
     s/n, ES-33007 Oviedo, Spain
    \label{OVIEDO}}
\titlefoot{Department of Physics, University of Oxford,
     Keble Road, Oxford OX1 3RH, UK
    \label{OXFORD}}
\titlefoot{Dipartimento di Fisica, Universit\`a di Padova and
     INFN, Via Marzolo 8, IT-35131 Padua, Italy
    \label{PADOVA}}
\titlefoot{Rutherford Appleton Laboratory, Chilton, Didcot
     OX11 OQX, UK
    \label{RAL}}
\titlefoot{Dipartimento di Fisica, Universit\`a di Roma II and
     INFN, Tor Vergata, IT-00173 Rome, Italy
    \label{ROMA2}}
\titlefoot{Dipartimento di Fisica, Universit\`a di Roma III and
     INFN, Via della Vasca Navale 84, IT-00146 Rome, Italy
    \label{ROMA3}}
\titlefoot{DAPNIA/Service de Physique des Particules,
     CEA-Saclay, FR-91191 Gif-sur-Yvette Cedex, France
    \label{SACLAY}}
\titlefoot{Instituto de Fisica de Cantabria (CSIC-UC), Avda.
     los Castros s/n, ES-39006 Santander, Spain
    \label{SANTANDER}}
\titlefoot{Dipartimento di Fisica, Universit\`a degli Studi di Roma
     La Sapienza, Piazzale Aldo Moro 2, IT-00185 Rome, Italy
    \label{SAPIENZA}}
\titlefoot{Inst. for High Energy Physics, Serpukov
     P.O. Box 35, Protvino, (Moscow Region), Russian Federation
    \label{SERPUKHOV}}
\titlefoot{J. Stefan Institute, Jamova 39, SI-1000 Ljubljana, Slovenia
     and Laboratory for Astroparticle Physics,\\
     \indent~~Nova Gorica Polytechnic, Kostanjeviska 16a, SI-5000 Nova Gorica, Slovenia, \\
     \indent~~and Department of Physics, University of Ljubljana,
     SI-1000 Ljubljana, Slovenia
    \label{SLOVENIJA}}
\titlefoot{Fysikum, Stockholm University,
     Box 6730, SE-113 85 Stockholm, Sweden
    \label{STOCKHOLM}}
\titlefoot{Dipartimento di Fisica Sperimentale, Universit\`a di
     Torino and INFN, Via P. Giuria 1, IT-10125 Turin, Italy
    \label{TORINO}}
\titlefoot{Dipartimento di Fisica, Universit\`a di Trieste and
     INFN, Via A. Valerio 2, IT-34127 Trieste, Italy \\
     \indent~~and Istituto di Fisica, Universit\`a di Udine,
     IT-33100 Udine, Italy
    \label{TU}}
\titlefoot{Univ. Federal do Rio de Janeiro, C.P. 68528
     Cidade Univ., Ilha do Fund\~ao
     BR-21945-970 Rio de Janeiro, Brazil
    \label{UFRJ}}
\titlefoot{Department of Radiation Sciences, University of
     Uppsala, P.O. Box 535, SE-751 21 Uppsala, Sweden
    \label{UPPSALA}}
\titlefoot{IFIC, Valencia-CSIC, and D.F.A.M.N., U. de Valencia,
     Avda. Dr. Moliner 50, ES-46100 Burjassot (Valencia), Spain
    \label{VALENCIA}}
\titlefoot{Institut f\"ur Hochenergiephysik, \"Osterr. Akad.
     d. Wissensch., Nikolsdorfergasse 18, AT-1050 Vienna, Austria
    \label{VIENNA}}
\titlefoot{Inst. Nuclear Studies and University of Warsaw, Ul.
     Hoza 69, PL-00681 Warsaw, Poland
    \label{WARSZAWA}}
\titlefoot{Fachbereich Physik, University of Wuppertal, Postfach
     100 127, DE-42097 Wuppertal, Germany
    \label{WUPPERTAL}}
\addtolength{\textheight}{-10mm}
\addtolength{\footskip}{5mm}
\clearpage
\headsep 30.0pt
\end{titlepage}
%
\pagenumbering{arabic} 
\setcounter{footnote}{0} %
\large
%
\section{Introduction}

\begin{figure}[b]
\begin{center}
\mbox{\epsfxsize\textwidth\epsfbox[50 100 550 550]{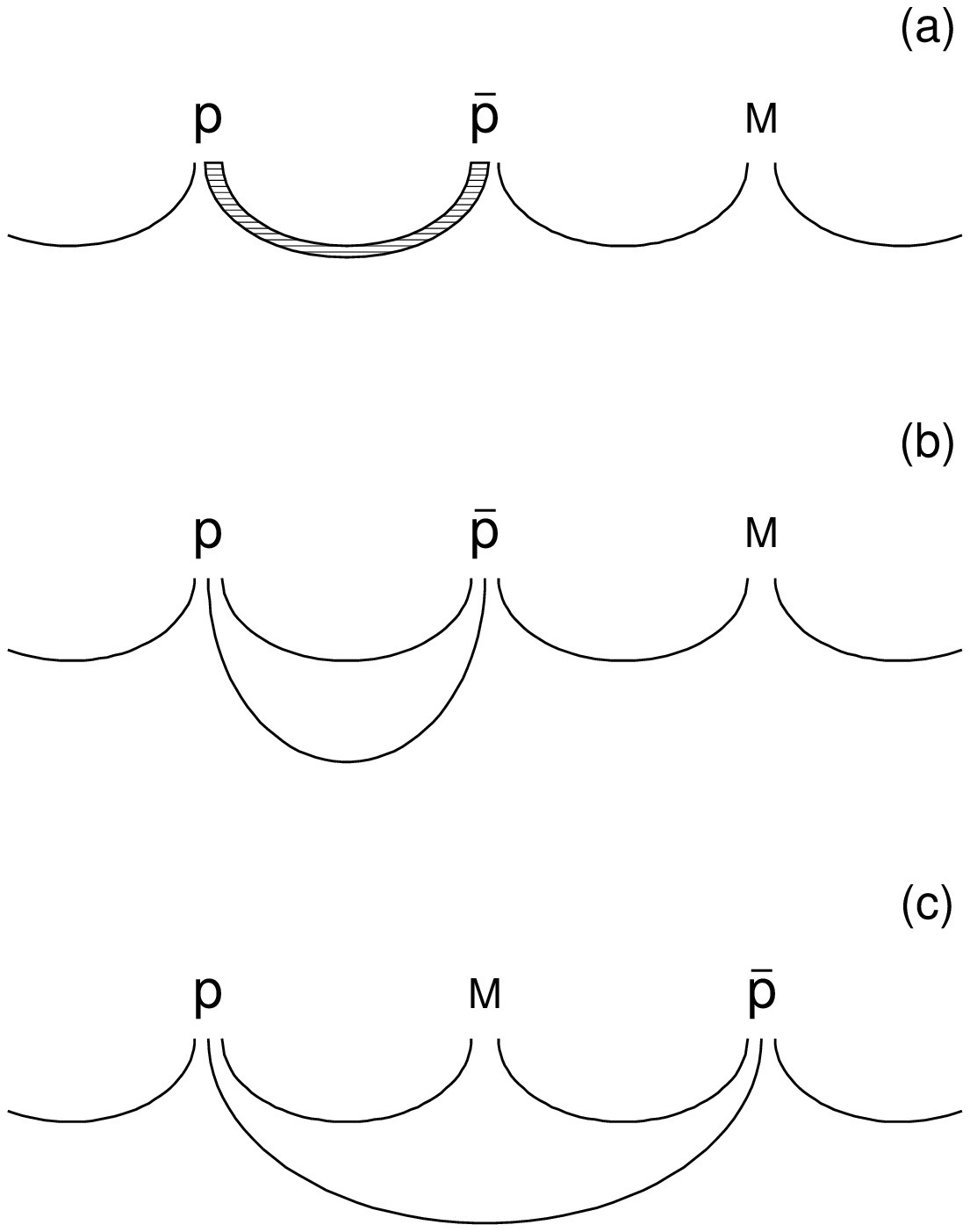}}
\end{center}
\caption[]{Illustration of \ppb production in the string model.
Each line represents a \qqb pair produced from potential energy
in the string.
(a) Production by a diquark-antidiquark pair (shown shaded) acting as
a fundamental unit.
(b) Through a step-wise process with two \qqb pairs forming an
effective diquark-antidiquark pair.
(c) Step-wise production with a mesonic state formed between the
\ppb pair (referred to as the popcorn effect).}
\label{fig:string}
\end{figure}

Baryon production from hadronic \zz decays, as interpreted in
string-fragmentation models, is pictured in Figure~\ref{fig:string}.
Hadronisation results from breaks in the string formed from the
colour-neutral system which stretches between the primary quarks
\cite{string}.
Breaks occur between virtual flavour-neutral \qqb pairs, with
mesons formed from string elements containing an adjacent $q$ and
$\bar{q}$.
Baryons are thought to be formed when breaks occur between
diquark-antidiquark pairs, the baryon being made from adjacent
diquark and quark \cite{diquark}.
A baryon and an antibaryon emerge as adjacent particles in rank along 
the string (`string-rank'), or possibly separated in rank with a
mesonic state between them.
Figure~\ref{fig:string}(a) represents the case where the diquark is
assumed to have a sufficiently large binding energy that it acts
like a fundamental unit.
Another possibility is to produce an `effective diquark' through a
step-wise process where two \qqb pairs are created, as shown in
Figure~\ref{fig:string}(b).
In this case a mesonic state also can be produced between the baryon
and antibaryon, seen in Figure~\ref{fig:string}(c).
This has been referred to as the `popcorn effect.'

In this paper, a novel method, using the rapidity-rank structure
of \ppb pairs, is used to study the mechanism of baryon production in
hadronic \zz decay.
A measurement of the relative frequency of occurrence of the rapidity-ordered
configuration $(i)$ \pmpbx, where $M$ is a charged meson, and $(ii)$ \ppb adjacent
in rapidity, is made to determine the magnitude of the popcorn effect.
This approach provides greater sensitivity than that used in previous studies
\cite{popcorn}.

\section{Data Sample and Event Selection}

This analysis is based on data collected with the DELPHI detector
\cite{delphi} at the CERN LEP collider in 1994 and 1995 at the \zz
centre-of-mass energy.
The charged-particle tracking information relies on three
cylindrical tracking detectors (Inner Detector, Time Projection
Chamber (TPC), and Outer Detector) all operating in a \mbox{1.2 T}
magnetic field.

The selection criteria for charged particles are: momentum
above 0.3 GeV/c, polar angle between $15\de$ and $165\de$, and
track length above 30 cm.
In addition, the impact parameters with respect to the beam axis
and along the longitudinal coordinate at the origin, are required to
be below 0.05 and 0.25 cm, respectively.
These impact parameter cuts decrease the number of protons which
result from secondary interactions in the detector. 
Also, protons from $\Lambda$ and $\Sigma$ decays are largely removed.

Hadronic events are selected by requiring at least three charged
particle tracks in each event hemisphere, and a total energy of all
charged particles exceeding 15 GeV.
The number of hadronic events is $\sim$ 2 million.

Charged particle identification is provided by a tagging procedure
which combines Cherenkov angle measurement from the RICH detector
with ionization energy loss measured in the TPC.
Details on the particle identification can be found in reference
\cite{delphi}.
In the present analysis, the combined-probability tag is required
to be at the `standard' level \cite{delphi}.
In addition, the polar angle for identified particles is restricted
to be in the barrel region, between $47\de$ and $133\de$.

\section{\bfm Rapidity-Rank Configurations \ppb and \pmpb}

This analysis studies \ppb correlations in the rapidity variable with
respect to the `thrust' direction.
The thrust direction approximates the directions of the primary $q$ and
$\bar{q}$, especially for two-jet events.
The rapidity, $y$, of a given particle is defined as
${1\over2} \, {{\ln}}\bigl((E+p_L) / (E-p_L) \bigr)$, where $p_L$ is
the momentum component parallel to the thrust axis, and $E$ is
the energy calculated using the particle mass as determined from RICH
and the measured momentum.
The restriction is made that events have only `one \p and one \bp'
in a given hemisphere.
Hemispheres are defined, one for positive $y$ and one for negative
$y$, with respect to the thrust direction.
Each hemisphere is considered independently.
The number of events with this selection is 27.6 thousand.
The background to this event sample can be determined from the number
of events that have two \ps or two \bps in a given hemisphere.
These events, 10.1 thousand, result mainly from \p or \bp misidentifications
and also from non-correlated baryon-antibaryon pairs (i.e., a \p and \bp from
different \BBb pairs).
This yields a 63\% purity, (27.6k$-$10.1k)/27.6k, for the \ppb sample.

A study of events from Jetset 7.3 \cite{jetset}, including detector simulation,
determined the \ppb pair detection efficiency to be $\sim$ 35\%, and the \ppb
pair purity to be $\sim$ 60\%, consistent with the above-mentioned value.
These values are nearly constant over the range of the analysis variable
\dymin defined later.
The efficiency is computed from the ratio of Jetset \ppb pairs detected,
to the total number of \ppb pairs generated.
The purity is obtained from the ratio of Jetset \ppb pairs detected and
congruous with a generated \ppb pair, to the total number of \ppb pairs
detected.

\begin{figure}[b]
\begin{center}
\mbox{\epsfxsize\textwidth\epsfbox[50 100 550 550]{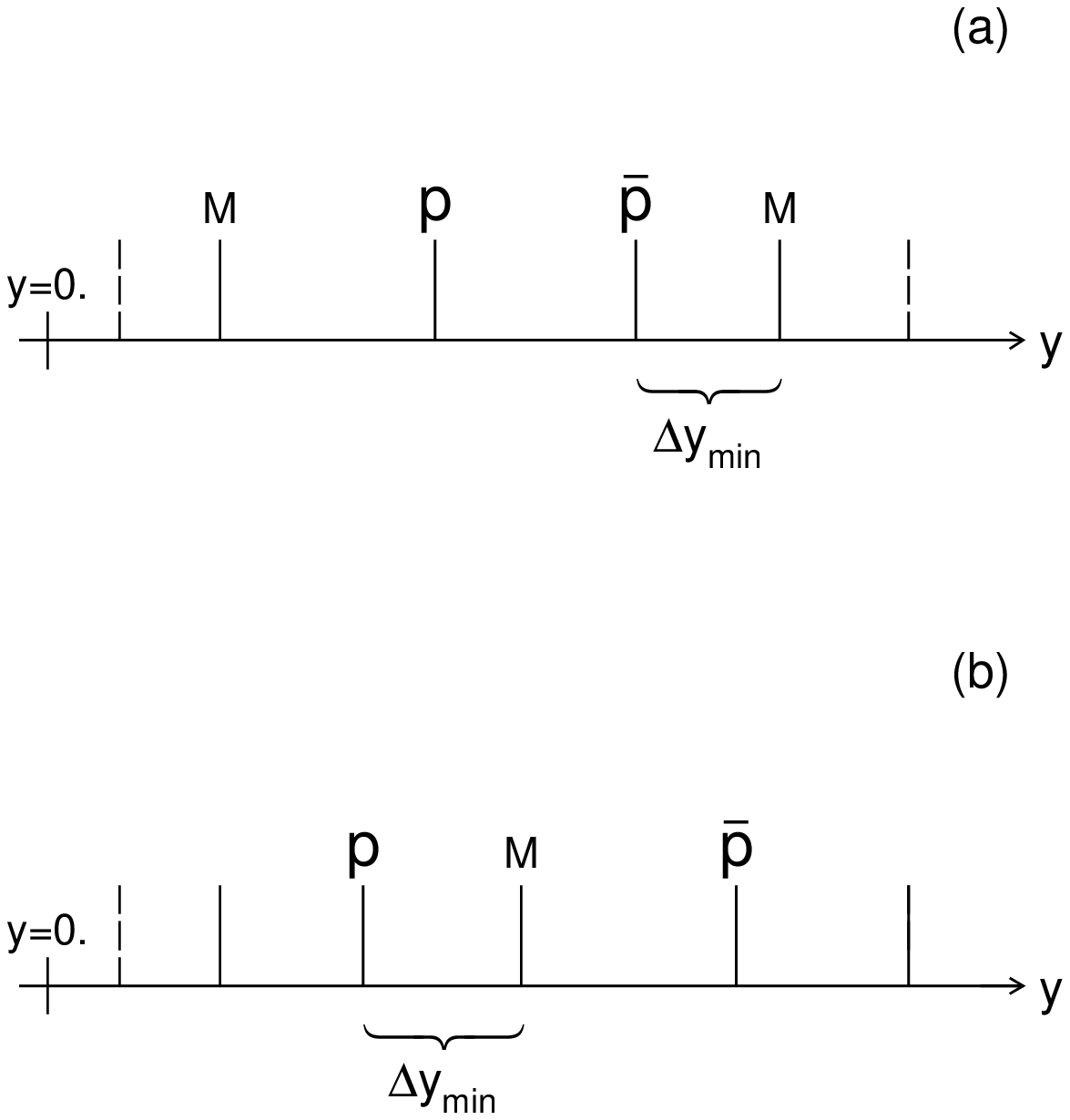}}
\end{center}
\caption[]{(a) An event hemisphere configuration with \p and \bp
adjacent in rapidity.
The rapidity-gap, $\dyminx$, indicates the distance to
the nearest particle external to the \ppb pair.
(b) An event configuration with a particle, $M$, between the
\p and \bp.
The rapidity-gap, $\dyminx$, denotes the distance of
the particle, $M$, to the nearest of \p or \bp.}
\label{fig:event}
\end{figure}

The charged particles in each event are ordered according to their
rapidity values as defined above.
The rapidity-rank is defined as the position that a particle has in
the rapidity chain.
In the following, two types of rapidity-rank configurations for \ppb
pairs are considered, and are shown in Figure~\ref{fig:event}.
The first is when the \p and \bp are adjacent in rapidity
(ranks differ by one unit).
The second is when the \p and \bp have one or more mesons between them.
The number of mesons is restricted to be at most three
(the ranks differ by two to four units).
This reduces the probability that the \p and \bp may have come from
different baryon-antibaryon pairs.
It should be noted that the rapidity configurations only approximately
portray the string-rank patterns as shown in Figure~\ref{fig:string}.
This is because of the softness of the fragmentation function and of
resonance decays which can mix the rapidity-ranks.

Since adjacent particles separated by a small rapidity gap have
a high probability to have `crossed-over' (reversed rank), this
study is performed as a function of the rapidity gap size.
For \ppb adjacent pairs, the concern is that a meson close in rapidity
to the \p or \bp may have crossed-over from an original
`string position' which was between the \ppb pair.
Correspondingly, for the \pmpb configuration, a meson on the outside of a
\ppb pair on the `string' may have crossed-over to be between the
\p and \bp in the rapidity variable.

To determine the relative amount of \ppb and \pmpb configurations in
the data the following ratio is calculated:
        $${\cal R}\bigl(\dyminx\bigr)=
  N(\pmpbx)\Big/\Bigl(N(\ppbx)+N(\pmpbx)\Bigr),  \eqno(1)$$
where $N(\ppbx)$ and $N(\pmpbx)$ represent the number of rapidity-rank
configurations of each type in the data sample, and are implicitly
a function of \dyminx, defined as follows.
For the \ppb case, Figure~\ref{fig:event}(a), \dymin is defined as
the absolute rapidity difference between the nearest adjacent
meson to either \p or \bp, whichever is smaller.
In the \pmpb case, Figure~\ref{fig:event}(b), \dymin is defined as
the absolute rapidity difference between either \p or \bp and the
particle in-between them, whichever is smaller.
If there is more than one particle in-between, then the particle which
is closest to being in the exact middle of the \ppb pair (and therefore
least likely to have crossed over) is the one considered.
With these definitions for \dyminx, the probability that a given rapidity
configuration will represent the actual rank order on the string
will be enhanced as \dymin is made larger.

If the production of \p and \bp are correlated, the rapidity gaps between
a \ppb pair are expected to be smaller than the gaps external to the pair.
In the present data the average size of rapidity gaps between the
\p and \bp for the \pmpb case (0.18 units) is $\sim$ 2/3 the size of
the adjacent rapidity gaps for the \ppb case (0.26 units).
To put the two cases on a more equal footing, the definition
of \dymin for the \ppb case includes a multiplicative factor of 2/3.
This is arbitrary but it provides for a better balance of the two
contributions when studying the ratio $\cal R$ over a range of \dyminx.
Excluded from the analysis are events where the particle with the
largest rank (i.e., smallest rapidity) is a \p or \bp.
This is to avoid the possibility that a low momentum particle may
not have been detected (or reconstructed) and could have formed a
small \dymin that was not considered.
The above treatments are applied to both data and model.

\begin{figure}[b]
\begin{center}
\mbox{\epsfxsize\textwidth\epsfbox[50 0 550 550]{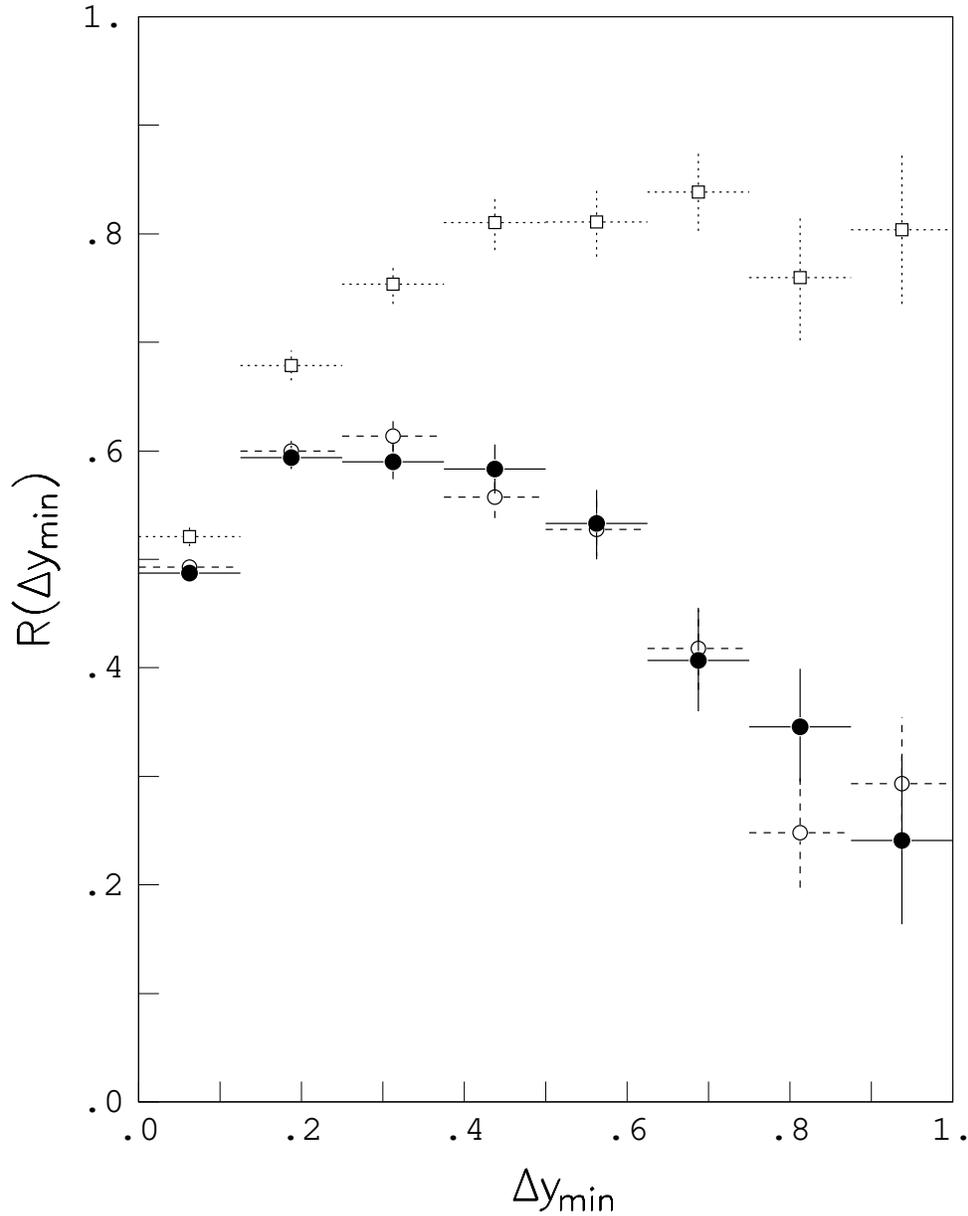}}
\end{center}
\caption[]{The relative amount, \Rdyx, of the \pmpb configuration
as a function of \dyminx.
The data points are indicated by solid circles.
The predictions from Jetset for two cases: no contribution from
the popcorn effect (open circles) and all popcorn effect (open
squares).}
\label{fig:ratio}
\end{figure}

The ratio \Rdy for the data is plotted in Figure~\ref{fig:ratio},
as solid circles.
Also shown are the predictions from Jetset 7.3 for the case when the
relative fraction of the \pmpb string-rank configuration is zero
and when it is 100\%, indicated by open circles and squares, respectively.
The errors are statistical.
A background subtraction of like-sign pairs ($p\,p$ and
$\bar{p}\,\bar{p}$) has been applied to the data and model to
remove contributions from uncorrelated baryon pairs and from
particle misidentifications.
The possible effect of variations (of order $\pm$ 30\%) in the fraction
of protons coming from resonances (like Deltas) was investigated, and
found to be negligible.
Standard DELPHI detector simulation along with charged particle
reconstruction and hadronic event selection are applied to the
events from Jetset 7.3 with parameters tuned as in reference \cite{tune}.
Since Jetset was run with a 50\% popcorn contribution, \ppb pairs
were separated into string-rank \pmpb (popcorn) and \ppb (non-popcorn)
components using the information on rank-order stored with the
Monte-Carlo events.
The prediction for the case with no contribution from \pmpb is seen to
fall for large \dyminx, as expected.
The case with 100\% contribution might be thought to rise to
the maximum value 1.0, but it flattens out possibly because
contributions, for example, from \ppb pairs with a $\pi^{0}('s)$ in
between become relatively more important for large \dyminx.
As seen in Figure~\ref{fig:ratio}, the data are consistent with
{\it no} contribution (or little) from \pmpb string configurations.
The $\chi^2$ between data and model was calculated as a function of 
the relative amount of \ppb and \pmpb configurations.
The $\chi^2$ is minimum over the range below 5\% popcorn
contribution; and, an upper-limit contribution of 15\% is determined
at 90\% confidence level.

\begin{figure}[b]
\begin{center}
\mbox{\epsfxsize\textwidth\epsfbox[50 0 550 550]{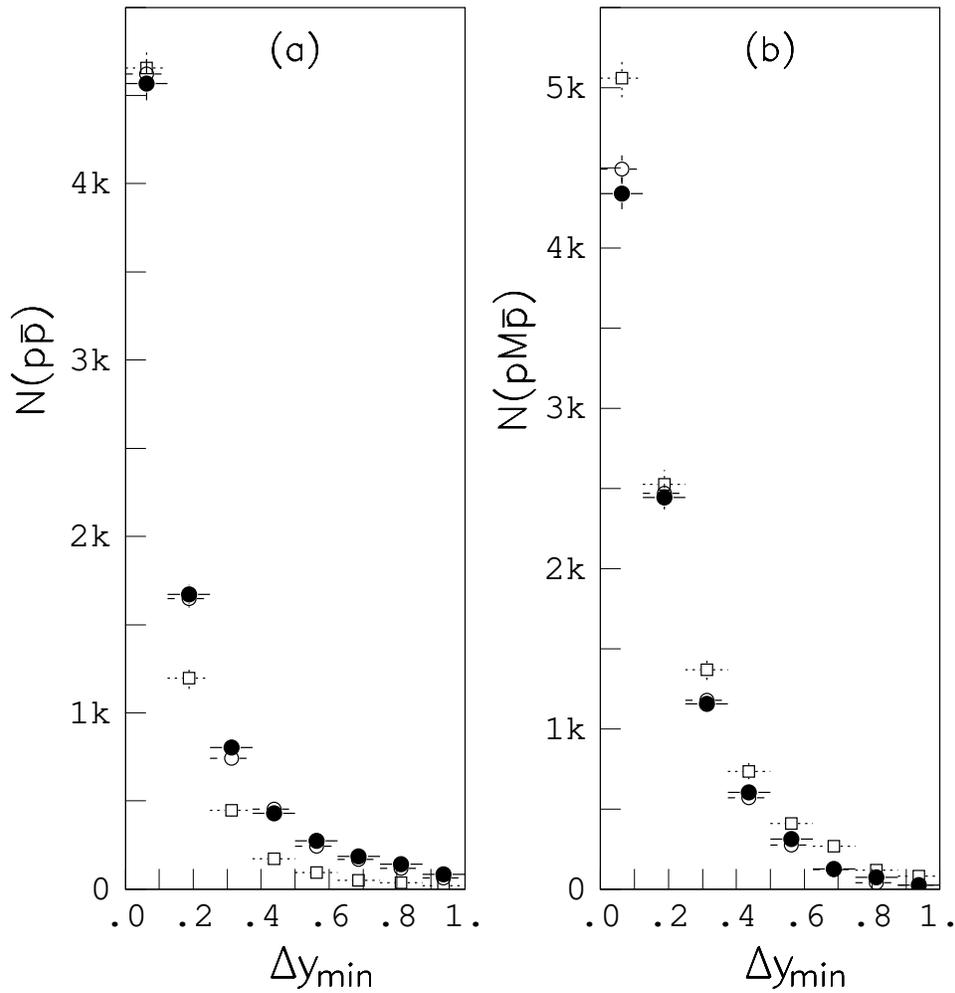}}
\end{center}
\caption[]{Distributions, $N(\ppbx)$ and $N(\pmpbx)$, of the number of
rapidity-rank configurations of each type as a function of \dyminx, in
(a) and (b), respectively.
The data points are indicated by solid circles.
The predictions from Jetset for two cases: no contribution from
the popcorn effect (open circles) and all popcorn effect (open squares).}
\label{fig:pnpp}
\end{figure}

For completeness, the distributions, $N(\ppbx)$ and $N(\pmpbx)$, of the
number of rapidity-rank configurations of each type as a function of
\dyminx, are displayed in Figure~\ref{fig:pnpp}.
The data are shown by the solid circles.
The predictions for the case when the relative fraction of the \pmpb
string-rank configuration is zero and when it is 100\%, are indicated
by open circles and squares, respectively.
In accord with the analysis above, consistency between the data and
the prediction for no-popcorn is evident for these distributions.

\section{\bfm The \ppb Rapidity Difference}

\begin{figure}[b]
\begin{center}
\mbox{\epsfxsize\textwidth\epsfbox[50 0 550 550]{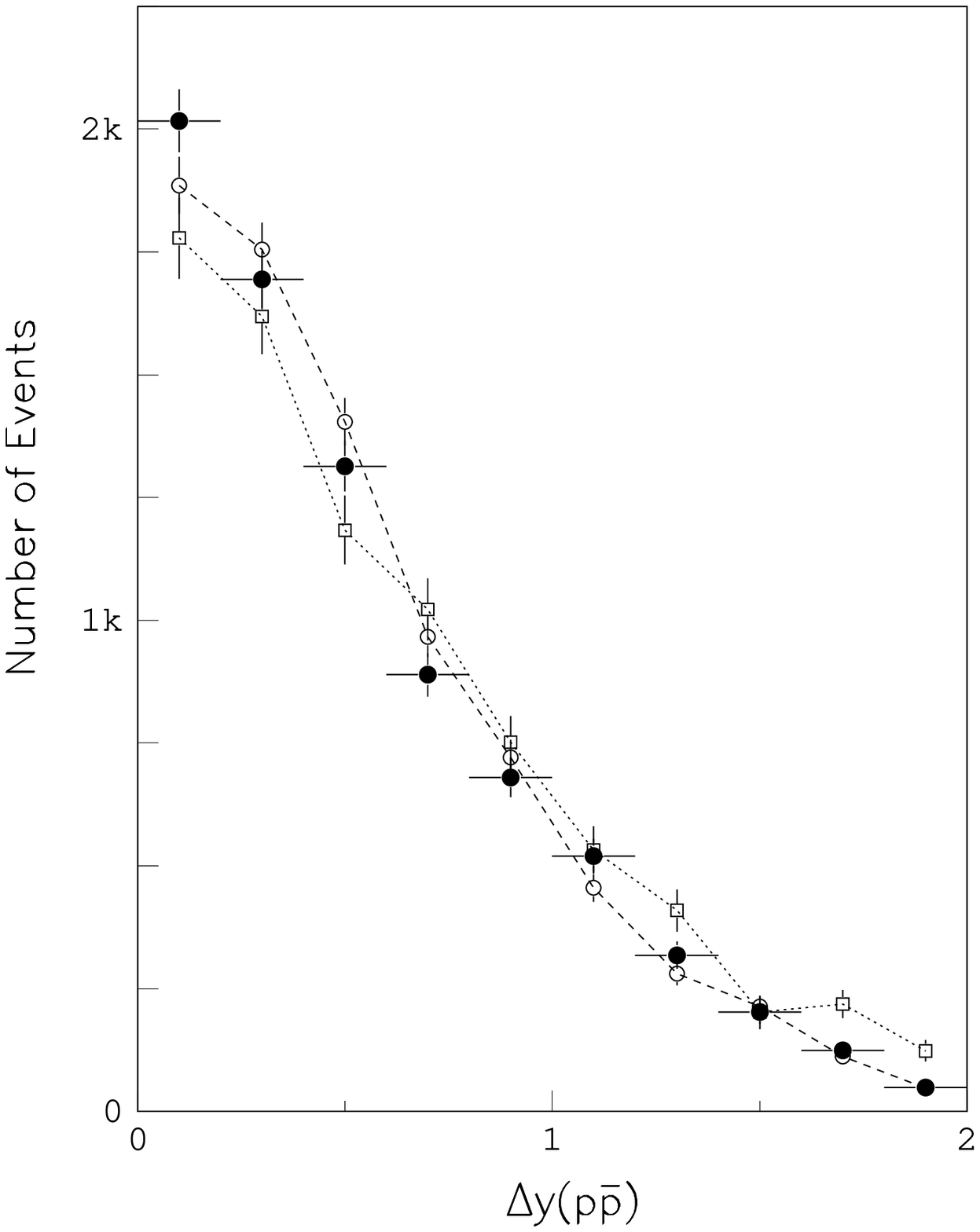}}
\end{center}
\caption[]{The distribution of \dyppb for the data (solid circles), and the
predictions of Jetset for no-contribution from the popcorn effect (open
circles), and for an all-popcorn effect (open squares).}
\label{fig:deltay}
\end{figure}

Previous studies of baryon-antibaryon (in particular, \llbx) rapidity
correlations have claimed evidence for the popcorn effect \cite{popcorn}.
In these studies, distributions of the \llb rapidity difference were compared
to predictions from the string-model Jetset.
To test the sensitivity of this method for the \ppb case, the
distribution of the \ppb rapidity difference, \dyppbx, for the data was
compared to the Jetset predictions for 100\% popcorn and for no-popcorn
contribution, as shown in Figure~\ref{fig:deltay}.
The thrust value was required to be greater than 0.96.
A background subtraction of like-sign pairs ($p\,p$ and
$\bar{p}\,\bar{p}$) has been applied to the data and model.
The all-popcorn assumption yields a mean value for \dyppb that is 11\%
larger than that for no-popcorn; without the thrust requirement the
difference is 5.5\%.
These values are in accordance with what is predicted in reference \cite{avedy}.
Even though this method is clearly not as sensitive as the one above-mentioned,
it can be seen that the data prefer the no-popcorn prediction.
The difference between the present result and that from \llb experiments might
indicate the importance of dynamical effects not incorporated in Jetset or simply
the inadequacy of the popcorn model, although no firm conclusion can be drawn yet.

\section{Conclusions}

The rapidity-rank structure of \ppb pairs was used to analyze the
mechanism of baryon production in hadronic \zz decay.
By comparing the relative occurrence in the data of the rapidity-ordered
configuration \pmpb, where $M$ is a meson, to that of \ppb adjacent pairs
with predictions from Jetset, it is found that the data can be explained
without requiring `string-ordered' \pmpb configurations.
The production of adjacent-rank \ppb pairs is sufficient to describe
the data.

\subsection*{Acknowledgements}
\vskip 3 mm
 We are greatly indebted to our technical 
collaborators, to the members of the CERN-SL Division for the excellent 
performance of the LEP collider, and to the funding agencies for their
support in building and operating the DELPHI detector.\\
We acknowledge in particular the support of \\
Austrian Federal Ministry of Science and Traffics, GZ 616.364/2-III/2a/98, \\
FNRS--FWO, Belgium,  \\
FINEP, CNPq, CAPES, FUJB and FAPERJ, Brazil, \\
Czech Ministry of Industry and Trade, GA CR 202/96/0450 and GA AVCR A1010521,\\
Danish Natural Research Council, \\
Commission of the European Communities (DG XII), \\
Direction des Sciences de la Mati$\grave{\mbox{\rm e}}$re, CEA, France, \\
Bundesministerium f$\ddot{\mbox{\rm u}}$r Bildung, Wissenschaft, Forschung 
und Technologie, Germany,\\
General Secretariat for Research and Technology, Greece, \\
National Science Foundation (NWO) and Foundation for Research on Matter (FOM),
The Netherlands, \\
Norwegian Research Council,  \\
State Committee for Scientific Research, Poland, 2P03B06015, 2P03B1116 and
SPUB/P03/178/98, \\
JNICT--Junta Nacional de Investiga\c{c}\~{a}o Cient\'{\i}fica 
e Tecnol$\acute{\mbox{\rm o}}$gica, Portugal, \\
Vedecka grantova agentura MS SR, Slovakia, Nr. 95/5195/134, \\
Ministry of Science and Technology of the Republic of Slovenia, \\
CICYT, Spain, AEN96--1661 and AEN96-1681,  \\
The Swedish Natural Science Research Council,      \\
Particle Physics and Astronomy Research Council, UK, \\
Department of Energy, USA, DE--FG02--94ER40817. \\

\clearpage

\clearpage

\clearpage

\clearpage

\clearpage


\newpage
\begin{thebibliography}{ref99}

\bibitem{string}
B. Andersson, G. Gustafson, G. Ingelman and T. Sj\"{o}strand,
Phys. Rep. {\bf 97} (1983) 31.

\bibitem{diquark}
A. Casher, H. Neuberger, and S. Nussinov,
Phys. Rev. {\bf D20} (1979) 179; \\
U.P. Sukhatme, K.E. Lassila and R. Orava,
Phys. Rev. {\bf D25} (1982) 2975; \\
T. Meyer,
Z. Phys. {\bf C12} (1982) 77; \\
B. Andersson, G. Gustafson, G. Ingelman and T. Sj\"{o}strand,
Z. Phys. {\bf C13} (1982) 361; \\
A. Bartl, H. Fraas and W. Majerotto,
Phys. Rev. {\bf D26} (1982) 1061; \\
A. Breakstone, et al.,
Z. Phys. {\bf C28} (1985) 335; \\
A. Breakstone, et al.,
Z. Phys. {\bf C36} (1987) 567; \\
M. Szczekowski,
Int. J. Mod. Phys. {\bf A4} (1989) 3985; \\
M. Anselmino, E. Predazzi, S. Ekelin, S. Fredriksson and D.B. Lichtenberg,
Rev. Mod. Phys. {\bf 65} (1993) 1199; \\
P. Ed\'{e}n and G. Gustafson,
Z. Phys. {\bf C75} (1997) 41.

\bibitem{popcorn}
H. Aihara, et al.,
Phys. Rev. Lett. {\bf 55} (1985) 1047; \\
OPAL Coll., P.D. Acton et al.,
Phys. Lett. {\bf B305} (1993) 415; \\
DELPHI Coll., P. Abreu et al.,
Phys. Lett. {\bf B318} (1993) 249; \\
ALEPH Coll., D. Buskulic et al.,
Z. Phys. {\bf C64} (1994) 361; \\
ALEPH Coll., D. Buskulic et al.,
Z. Phys. {\bf C71} (1996) 357; \\
DELPHI Coll., P. Abreu et al.,
Phys. Lett. {\bf B416} (1998) 247; \\
OPAL Coll., G. Abbiendi et al.,
CERN-EP/98-114, accepted by Eur. Phys. J. {\bf C}.
 
\bibitem{delphi}
DELPHI Coll., P. Aarnio et al.,
Nucl. Instr. and Meth. {\bf A303} (1991) 233; \\
DELPHI Coll., P. Abreu et al.,
Nucl. Instr. and Meth. {\bf A378} (1996) 57; \\
DELPHI Coll., P. Aarnio et al.,
Phys. Lett. {\bf B240} (1990) 271.

\bibitem{jetset}
T. Sj\"{o}strand and M. Bengtsson,
Comp. Phys. Comm. {\bf 43} (1987) 367; \\
T. Sj\"{o}strand, CERN-TH.6488/92, May 1992, Revised Sept. 1992.

\bibitem{tune}
DELPHI Coll., P. Abreu et al.,
Z. Phys. C73 (1996) 11.

\bibitem{avedy}
B. Anderson, G. Gustafson and T. Sj\"{o}strand,
Phys. Scripta {\bf 32} (1985) 574.

\end{thebibliography}
\end{document}